\documentclass{aip-cp}

\usepackage[numbers]{natbib}
\usepackage{rotating}
\usepackage{graphicx}

\begin{document}

\title{Quartetting Wave Function Approach to $^{20}$Ne:\\ Shell Model and Local Density Approximation
}

\author
{G. R\"{o}pke}
\noteref{note1}
\affil
{Institut f\"{u}r Physik, Universit\"{a}t Rostock, D-18051 Rostock, Germany}
\authornote[note1]
{e-mail: gerd.roepke@uni-rostock.de}

\maketitle

\begin{abstract}
We investigate $\alpha$-like correlations in $^{20}$Ne.
A quartet of nucleons (different spin/isospin) is moving in a mean field produced by the $^{16}$O core nucleus. 
Improving the Thomas-Fermi model (local density approach), a shell model is
considered for the core nucleus. The effective potential of the $\alpha$-like quartet and the wave function for the center-of-mass (c.o.m.) motion are calculated and compared with other approaches.
\end{abstract}

\section{INTRODUCTION}

The quasiparticle concept has been proven to be very successful in strongly interacting quantum systems, such as nuclear matter, see \cite{RingSchuck}.
Effective mean-field approaches are frequently used to describe the nuclear matter EOS and related quantities, but also for the shell model of nuclei.
Kinetic approaches such as the BUU equation are based on a single-particle description.

Correlations have to be included to improve the single-particle approach. A famous step in this direction is the treatment of pairing where a self-consistent mean-field approach is given within the BCS approach. Nowadays, pairing is a well established concept in nuclear physics.

A subsequent step is the inclusion of four-particle correlations. 
The $\alpha$ particle is strongly bound (7.1 MeV/$A$ in contrast  to 1.1 MeV/$A$ for the deuteron)
so that quartets consisting of four nucleons $\{ n_\uparrow, n_\downarrow, p_\uparrow, p_\downarrow \}$ 
are important structures. For instance, signatures of $\alpha$-like quartetting are found in matter 
at low densities. They are relevant for the nuclear matter equation of state \cite{R}, 
the Hoyle state \cite{THSR}, and the $\alpha$ decay of heavy nuclei, see \cite{Po,Xu,Xu1}.

Different approaches have been used to describe quartetting in matter, see \cite{Toh17} for references. 
We discuss the quartetting wave function approach recently worked out to describe the $\alpha$ decay 
of heavy nuclei \cite{Po}. The basic idea is to investigate the center-of mass (c.o.m.) motion 
of the quartet treating the surrounding matter as mean field.
As for the single particle and pairing cases, the nucleons of the quartet under consideration have 
to obey the Pauli exclusion principle. 
However, any correlations with the environment are neglected.

This concept was demonstrated within a local-density approximation (Thomas-Fermi model) and has been applied to $^{208}$Po 
\cite{Po,Xu} and further heavy, $\alpha$-emitting nuclei \cite{Xu1}. 
The local-density approximation is too rough to reproduce details, in particular the signatures of shell structure of the core nucleus.
The introduction of shell structure to be discussed in this work is a consequent next step to improve the local-density approximation.

A striking result of the quartetting wave function approach is a nearly constant effective potential for the quartet c.o.m. motion inside the core nucleus.
This result, which is in contradiction to other approaches to define a cluster-cluster effective potential, was investigated within a
model considering non-interacting particles in an external, harmonic oscillator potential \cite{wir} confirming the results of Thomas-Fermi calculations. 
It has been shown that also in this model case, the effective potential for the quartet c.o.m. motion is nearly constant inside the nucleus.

Here, we continue to work out this approach \cite{wir}. We consider  the nucleus $^{20}$Ne where a quartet is moving on top of
the $^{16}$O core nucleus. We compare our effective quartet potential with optical potentials describing 
the elastic $^{16}$O $- \alpha$ scattering. In addition, instead of the harmonic oscillator model we consider 
more realistic shell-model states, obtained from a Woods-Saxon potential. Finally, the effect of Pauli blocking will be discussed.

The work is in tight relation to the THSR (Tohsaki-Horiuchi-Schuck-R\"opke) approach which was also applied to $^{20}$Ne \cite{Bo,Bo2,Bo3}. There, quartetting correlations are also considered in the nuclear medium. We discuss the THSR approach at the end of this work.

\section{THE QUARTET WAVE EQUATION }


The treatment of the interacting many-nucleon system needs some approximations 
which may be obtained in a consistent way from a Green's functions approach. 
The quartetting wave function approach \cite{Po,wir} considers the wave function 
$\Psi({\bf r}_1 {\bf r}_2 {\bf r}_3{\bf r}_4)$ of the quartet 
(spin and isospin variables are dropped) which obeys the in-medium wave equation
\begin{eqnarray}
&&[E_4\!-\!\hat h_1\! -\!\hat h_2\!-\! \hat h_3\! - \!\hat h_4]\Psi({\bf r}_1 {\bf r}_2 {\bf r}_3{\bf r}_4)\!=\!\!\!
\int \!\! d^3 {\bf r}_1'\,d^3 {\bf r}_2' \langle {\bf r}_1{\bf r}_2|\hat B(1,2) \,\,\hat V_{N-N}| {\bf r}_1'{\bf r}_2'\rangle
\Psi({\bf r}_1'{\bf r}_2'{\bf r}_3{\bf r}_4)\nonumber \\ && +
\int d^3 {\bf r}_1'\,\,d^3{\bf r}_3'  \langle {\bf r}_1{\bf r}_3|\hat B (1,3) \,\,\hat V_{N-N}|
{\bf r}_1'{\bf r}_3'\rangle \Psi({\bf r}_1'{\bf r}_2{\bf r}_3'{\bf r}_4)
+ {\rm four\,\, further \,\,permutations,}
\label{15}
\end{eqnarray}
with the single-quasiparticle Hamiltonian (single-nucleon shell states $|n,\nu \rangle$)
\begin{equation}
\label{spPauli}
 \hat h_i=\frac{\hbar^2 \hat p_i^2}{2m}+ [1 - \hat f_{\nu_i}]\, V_{\nu_i}^{\rm mf}(\hat r),\qquad
\hat f_\nu =\sum_n^{{\rm occ.}}| n,\nu \rangle \langle n,\nu |
\end{equation}
indicates the phase space that according to the Pauli principle is not available for an interaction process of a nucleon 
with internal quantum state $\nu=\sigma,\,\tau$.
The six  nucleon-nucleon interaction terms contain besides the nucleon-nucleon potential $\hat V_{N-N}$ also the 
 blocking operator $ \hat B(1,2)=[1-\hat f_1-\hat f_2]$ for the first term on the r.h.s. of Eq. (\ref{15}), ect. The full interaction $\hat V_{N-N}$ acts within the $\alpha$-like cluster. 
The mean-field potential $V_{\nu_i}^{\rm mf}(\hat r)$ contains the Coulomb potential as well as the nucleon-nucleon interaction 
$V^{\rm ext}( r)$ of the core nucleus. The Pauli blocking terms are not easily treated 
as discussed in the following.

A main aspect of the cluster approach is the introduction of the center-of-mass (c.o.m.) motion $\bf R$ 
as new collective degree of freedom, and ${\bf s}_j=\{\bf S,s,s'\}$ for the intrinsic motion
(Jacobi-Moshinsky coordinates).
As shown in \cite{Po}, the normalized quartet wave function $\Phi({\bf R},{\bf s}_j) $,
\begin{equation}
\int d^3R\,\int d^9s_j\,|\Phi({\bf R},{\bf s}_j)|^2 =1,
\end{equation}
can be decomposed in a unique way (up to a phase factor),
\begin{equation}
\label{4}
\Phi({\bf R},{\bf s}_j)=\varphi^{{\rm intr}}({\bf s}_j,{\bf R})\,\psi({\bf R})
\end{equation}
with the individual normalizations 
 \begin{equation}
 \label{normS}
\int d^3R\,|\psi({\bf R})|^2=1
\,, \qquad {\rm and} 
\int d^9s_j |\varphi^{{\rm intr}}({\bf s}_j,{\bf R})|^2=1
\end{equation}
for arbitrary ${\bf R}$.

The Hamiltonian of a cluster 
may be written as 
\begin{equation}
H=\left(-\frac{\hbar^2}{8m} \nabla_R^2+T[\nabla_{s_j}]\right)\delta^3({\bf R}-{\bf R}')\delta^3({\bf s}_j-{\bf s}'_j)
+V({\bf R},{\bf s}_j;{\bf R}',{\bf s}'_j)
\end{equation}
with the kinetic energy of the c.o.m. motion and the kinetic energy of the internal motion of the cluster, $T[\nabla_{s_j}]$. 
The interaction $V({\bf R},{\bf s}_j;{\bf R}',{\bf s}'_j)$  contains the mutual interaction $V_{ij}({\bf r}_i,{\bf r}_j,{\bf r}'_i,{\bf r}'_j)$ 
between the particles 
as well as the 
interaction with an external potential (for instance, the potential of the core nucleus). 

For the c.o.m. motion we have the wave equation
\begin{eqnarray}
\label{9}
&&-\frac{\hbar^2}{8m} \nabla_R^2\psi({\bf R})-\frac{\hbar^2}{Am}\int d^9s_j \varphi^{{\rm intr},*}({\bf s}_j,{\bf R}) 
[\nabla_R \varphi^{{\rm intr}}({\bf s}_j,{\bf R})][\nabla_R\psi({\bf R})]-
\\ &&
-\frac{\hbar^2}{8m}\int\!\! d^9s_j \varphi^{{\rm intr},*}({\bf s}_j,{\bf R}) 
[ \nabla_R^2 \varphi^{{\rm intr}}({\bf s}_j,{\bf R})] \psi({\bf R})
+\!\!\int \!\! d^3R'\,W({\bf R},{\bf R}')  \psi({\bf R}')\!=\!E\,\psi({\bf R})\,\nonumber 
\end{eqnarray}
with the c.o.m. potential
\begin{eqnarray}
\label{9c}
W({\bf R},{\bf R}')&=&\int d^9s_j\,d^9s'_j\,\varphi^{{\rm intr},*}({\bf s}_j,{\bf R}) \left[T[\nabla_{s_j}]
\delta^3({\bf R}-{\bf R}')\delta^9({\bf s}_j-{\bf s}'_j)\right.\nonumber \\&&\left.
+V({\bf R},{\bf s}_j;{\bf R}',{\bf s}'_j)\right]
\varphi^{{\rm intr}}({\bf s}'_j,{\bf R}')\,.
\end{eqnarray}
For the intrinsic motion we find the wave equation
\begin{eqnarray}
\label{10}
&&-\frac{\hbar^2}{4m}  \psi^*({\bf R}) [\nabla_R\psi({\bf R})]
[\nabla_R \varphi^{{\rm intr}}({\bf s}_j,{\bf R})]
-\frac{\hbar^2}{8m}  |\psi({\bf R})|^2
\nabla_R^2 \varphi^{{\rm intr}}({\bf s}_j,{\bf R})
\nonumber \\ &&
+\int d^3R'\,d^9s'_j\, \psi^*({\bf R}) \left[T[\nabla_{s_j}]
\delta^3({\bf R}-{\bf R}')\delta^9({\bf s}_j-{\bf s}'_j)\right.\nonumber \\&& \left.
+V({\bf R},{\bf s}_j;{\bf R}',{\bf s}'_j)\right]
 \psi({\bf R}')\varphi^{{\rm intr}}({\bf s}'_j,{\bf R}')=F({\bf R}) \varphi^{{\rm intr}}({\bf s}_j,{\bf R})\,.
\end{eqnarray}
The respective c.o.m. and intrinsic Schr\"odinger
equations are coupled by contributions containing the expression
$\nabla_R \varphi^{{\rm intr}}({\bf s}_j,{\bf R})$ which will be
neglected in the present work. This expression vanishes in homogeneous matter.
No investigations of such gradient terms have been performed yet for inhomogeneous systems.

\section{QUARTETS IN NUCLEI IN THOMAS-FERMI APPROXIMATION}

We emphasize that we should allow for non-local interactions. In particular, the Pauli blocking considered below
is non-local. Also the nucleon-nucleon interaction can be taken as non-local potential. 
To simplify the calculations, often local approximations are used,
\begin{equation}
W({\bf R},{\bf R}')\approx W({\bf R})\delta^3({\bf R}-{\bf R}'),\qquad W({\bf R})=W^{\rm ext}({\bf R})+W^{\rm intr}({\bf R}).
\end{equation}
$W^{\rm ext}({\bf R})=W^{\rm mf}({\bf R})$ is the contribution of external 
potentials, here the mean field of the core nucleons.
The interaction within the cluster according  Eq. (\ref{10}) gives the contribution $W^{\rm intr}({\bf R})$.

Having the nucleon densities of the core nucleus to our disposal, the mean fields are easily calculated.
The mean-field contribution $W^{\rm mf}({\bf R})$ is obtained by double folding the density distribution of the  core 
nucleus and the intrinsic density distribution of the quartet at c.o.m. position $\bf R$ with the interaction potential. 
For the bound quartet, an  $\alpha$-like Gaussian density distribution has been taken. 

For the Coulomb interaction we calculate the double-folding potential
\begin{equation}
\label{VCoul}
V^{\rm Coul}_{\alpha - {\rm O}}( R) = \int d^3 r_1 \int d^3 r_2 \rho_{{\rm O}} ({\bf r}_1) \rho_\alpha ({\bf r}_2) 
\frac{e^2}{|{\bf R}-{\bf r}_1+{\bf r}_2|}\,.
\end{equation}
The charge density of the $\alpha$ nucleus according to 
\begin{equation}
\label{nqalpha}
 \rho_\alpha( r)=0.21144\,\,{\rm fm}^{-3} \,e^{-0.7024\,\, r^2/{\rm fm}^2} 
\end{equation}
reproduces the measured rms point radius 1.45 fm. For the density distribution
of $^{16}$O, the expression \cite{Qu2011}
\begin{equation}
\label{Qu}
n^{\rm WS}_{B,{\rm O}}( r) = \frac{0.168 \,{\rm fm}^{-3}}{1+e^{(r/{\rm fm}-2.6)/0.45}}
\end{equation}
was given which reproduces the rms point radius 2.6201 fm, or Gaussians \cite{wir}.
The convolution integral (\ref{VCoul}) is easily evaluated in Fourier representation and gives 
for the parameter values considered here \cite{wir}
\begin{eqnarray}
\label{coulao}
&& V^{\rm Coul}_{\alpha - {\rm O}}( R)=\frac{16 \times 1.44}{R} {\rm MeV\,\, fm} \nonumber \\&& \times\left[{\rm Erf}(0.76829\,\, R/{\rm fm})-0.9097 \,\,(R/{\rm fm})\,\,e^{-0.22736\,\,R^2/{\rm fm}^2}\right]\,.
\end{eqnarray}

For the nucleon-nucleon contribution to the mean field, 
a parametrized effective nucleon interaction (distance $s$)
\begin{equation}
\label{VNN}
 V_{N-N}(s/{\rm fm})=c\, \exp(-4 s)/(4 s)-d\, \exp(-2.5 s)/(2.5 s)
\end{equation}
can be used which is motivated by the M3Y interaction \cite{M3YReview}, 
$s$ denotes the distance of nucleons. The parameters $c, d$ are adapted 
to reproduce known data, see \cite{Po,Xu,Xu1}
for the case of a lead core nucleus. For the oxygen core nucleus,
parameter values $c,d$ are given below in Eq. (\ref{cd}).
As also known from other mean-field approaches, we fit the mean field parameter to measured data.
The nucleonic contribution $V^{\rm N-N}_{\alpha - {\rm O}}( R)$ to the mean field is calculated in analogy to Eq. (\ref{VCoul}) replacing the Coulomb interaction by the nucleon interaction (\ref{VNN}).
With both contributions, the mean-field part of the cluster potential is 
\begin{equation}
\label{ext}
W^{\rm ext}({\bf R})=W^{\rm mf}({\bf R})=V^{\rm Coul}_{\alpha - {\rm O}}( R)+V^{\rm N-N}_{\alpha - {\rm O}}( R). 
\end{equation}

The local approximation $W^{\rm intr}({\bf R})$ for the intrinsic contribution 
to the effective c.o.m. potential is more involved. 
It contains the binding energy of the cluster taking into account the Pauli blocking 
of the surrounding matter. The local density approximation neglects any gradient terms 
so that the results for homogeneous matter can be used.

The intrinsic wave equation (\ref{10})  describes in the zero density limit
the formation of an $\alpha$ particle with binding energy $B_\alpha= 28.3$
MeV. For homogeneous matter, the binding energy will be reduced
because of Pauli blocking. The shift of the binding energy is determined by the baryon
density $n_B=n_n+n_p$, For the c.o.m. momentum ${\bf P}=0$, the Pauli
blocking term depends on the  baryon density $n_B$ \cite{Po,wir} as
\begin{equation}
\label{WPauli}
 W^{\rm Pauli}(n_B)\approx 4515.9\, {\rm MeV\, fm}^3 n_B -100935\, {\rm MeV\, fm}^6 n_B^2+1202538\, {\rm MeV\, fm}^9 n_B^3\,.
\end{equation}
This fit formula is valid in the density  
region $n_B \le 0.03$ fm$^{-3}$.
In particular, the bound state is dissolved and merges with the continuum
of scattering states at the critical density $n_{\rm crit}= 0.02917$ fm$^{-3}$ 
(introduced as Mott density).
For the intrinsic wave function of the quartet we can assume an $\alpha$-like Gaussian
to describe the bound state. The width parameter of the free $\alpha$ particle is only 
weakly changed when approaching the critical density, see Ref. \cite{Po}.

Below  the critical density, $n_B \le n_{\rm crit}$, the intrinsic potential
\begin{equation}
\label{WeffR}
W^{\rm intr}({\bf R})=-B_\alpha+ W^{\rm Pauli}[n_B({\bf R})], \qquad n_B \le n_{\rm crit}
\end{equation}
results in local density approximation. The intrinsic energy of the quartet for densities above the critical one is a minimum if all four nucleons are at the Fermi energy  (ideal Fermi gas),
for symmetric matter and  $n_B \ge n_{\rm crit}$
\begin{equation}
W^{\rm intr}({\bf R})=4 E_F[n_B({\bf R})], \qquad E_F(n_B)=(\hbar^2/2m) (3 \pi^2n_B/2)^{2/3}.
\end{equation}

\section{THOMAS-FERMI RULE AND RESULTS FOR  $^{20}$Ne IN LOCAL DENSITY APPROXIMATION}
\label{TFR}

The quartetting wave function approach to  $^{20}$Ne in local density approximation 
has been considered in \cite{wir}. We repeat results for the effective potential $W( {\bf R})$
and the wave function $\psi({\bf R})$. Instead of a microscopic approach using a nucleon-nucleon interaction, we use empirical data to construct these quantities.

The mean-field contribution $W^{\rm ext}( R)$ (\ref{ext})
is given by the double-folding Coulomb and $N-N$ potentials. 
Empirical values for the densities of the $\alpha$ particle (\ref{nqalpha}) 
and the $^{16}$O core nucleus (\ref{Qu})
are known from scattering experiments (rms radii) so that the Coulomb interaction 
$V^{\rm Coul}_{\alpha - {\rm O}}( R)$ (\ref{coulao}) as well as the  nucleon-nucleon interaction 
$V^{\rm N-N}_{\alpha - {\rm O}}( R)$ can be calculated. 

With respect to $W^{\rm intr}({\bf R})$, the local density approximation is 
also very simple:\\
There are two regions separated by the critical radius $r_{\rm crit} =3.302$ fm where the density 
of the $^{16}$O core nucleus (\ref{Qu}) has the critical value 
$n_B(r_{\rm crit})= n_{\rm crit}= 0.02917$ fm$^{-3}$. 
We obtain $-B_\alpha+W^{\rm Pauli}[n_B(r_{\rm crit})]=4 E_F[n_B(r_{\rm crit})]$, and the bound state merges 
with the continuum of scattering states. 

For $R > r_{\rm crit}$, the intrinsic part $W^{\rm intr}( R)$ 
contains the bound state energy -28.3 MeV of the free $\alpha$ particle which is shifted because of Pauli blocking. 
At $r_{\rm crit}$, the bound state merges with the continuum so that we have the condition (symmetric matter)
\begin{equation}
 W(r_{\rm crit})= W^{\rm ext}(r_{\rm crit})+4 E_F(n_{\rm crit}) =\mu_4,
\end{equation}
the intrinsic wave function changes from a bound state case to four uncorrelated quasiparticles on top of the Fermi sphere (the states below the Fermi energy are already occupied).

For $R < r_{\rm crit}$, in addition to the mean-field contribution $W^{\rm ext}( R)$ the Fermi energy 
$4 E_F[n( R)]$ appears. Within the Thomas-Fermi model, for a given potential $W^{\rm ext}( R)$ 
the density is determined by the condition that $W^{\rm ext}(R )+4 E_F[n_B( R)]$ remains a constant, 
here $\mu_4$.
We find the effective potential $W^{\rm TF}( R)$ which is continuous but has a kink at $r_{\rm crit}$.
It is an advantage of the Thomas-Fermi model that the condition $W^{\rm TF}( R)=\mu_4=$ const holds 
for the entire region $R < r_{\rm crit}$, 
independent of the mean-field potential $W^{\rm ext}( R)$ and the corresponding density distribution. 
We analyze this property in the following section.

Whereas the Coulomb part to the external potential as well as the intrinsic part of the effective potential 
$W^{\rm TF}( R)$ are fixed, both parameters $c,d$ for the $N-N$ part of the external potential 
can be adjusted such 
that measured data are reproduced. In the case of heavy nuclei which are $\alpha$ emitters like $^{212}$Po \cite{Po}, we have to formulate two conditions:\\ 
i) For $\alpha$ emitters, the solution of the c.o.m. wave equation (neglecting decay) gives the energy eigenvalue $E_{\rm tunnel}$. 
This eigenvalue should coincide with the measured energy after decay as given by the $Q$ value.\\
ii) This value $E_{\rm tunnel}$ should coincide with the value $\mu_4$. 
Within the local density approach, this is the value the four nucleons must have to implement them into the core nucleus. We denote this condition $E_{\rm tunnel}=\mu_4$ as the Thomas-Fermi rule \cite{wir}.
With both conditions, the parameter $c,d$ for the double folding $N - N$ interaction potential are found, 
and values for the preformation factor and the half life of the $\alpha$ decay have been obtained in heavy nuclei, see Ref. \cite{Po,Xu,Xu1} where further discussions have been given.

In contrast to the $\alpha$ decay of $^{212}$Po where the $Q$ value can be used to estimate the chemical potential $\mu_4$ \cite{Po}, the  $^{20}$Ne is stable. However, 
we can use the additional binding when going from  $^{16}$O ($B_{^{16}{\rm O}}=127.66$ MeV) to  $^{20}$Ne ($B_{^{20}{\rm Ne}}=160.645$ MeV) 
adding the four nucleons. The difference fixes  
the position of the in-core effective potential $\mu_4 =B_{^{16}{\rm O}}-B_{^{20}{\rm Ne}}=-33.0$ MeV. 

Another condition is that  the solution of the Schr{\"o}dinger equation for the four-nucleon c.o.m. motion in the effective potential $W( R)$ gives the energy eigenvalue 
$E_{\alpha, {\rm bound}}$ at this value -33 MeV so that the $\alpha$-like cluster is at the 
Fermi energy $\mu_4 $ (see also the discussion in Ref. \cite{Xu}). 
Both conditions are used  to fix the parameters $c, d$.  The values 
\begin{equation}
\label{cd}
 c=4650\,\,{\rm  MeV \,\,\,\, and}\,\,\, d=1900\,\,{\rm MeV} 
\end{equation}
have been found \cite{wir}.

\begin{figure}[h]
  \centerline{\includegraphics[width=250pt]{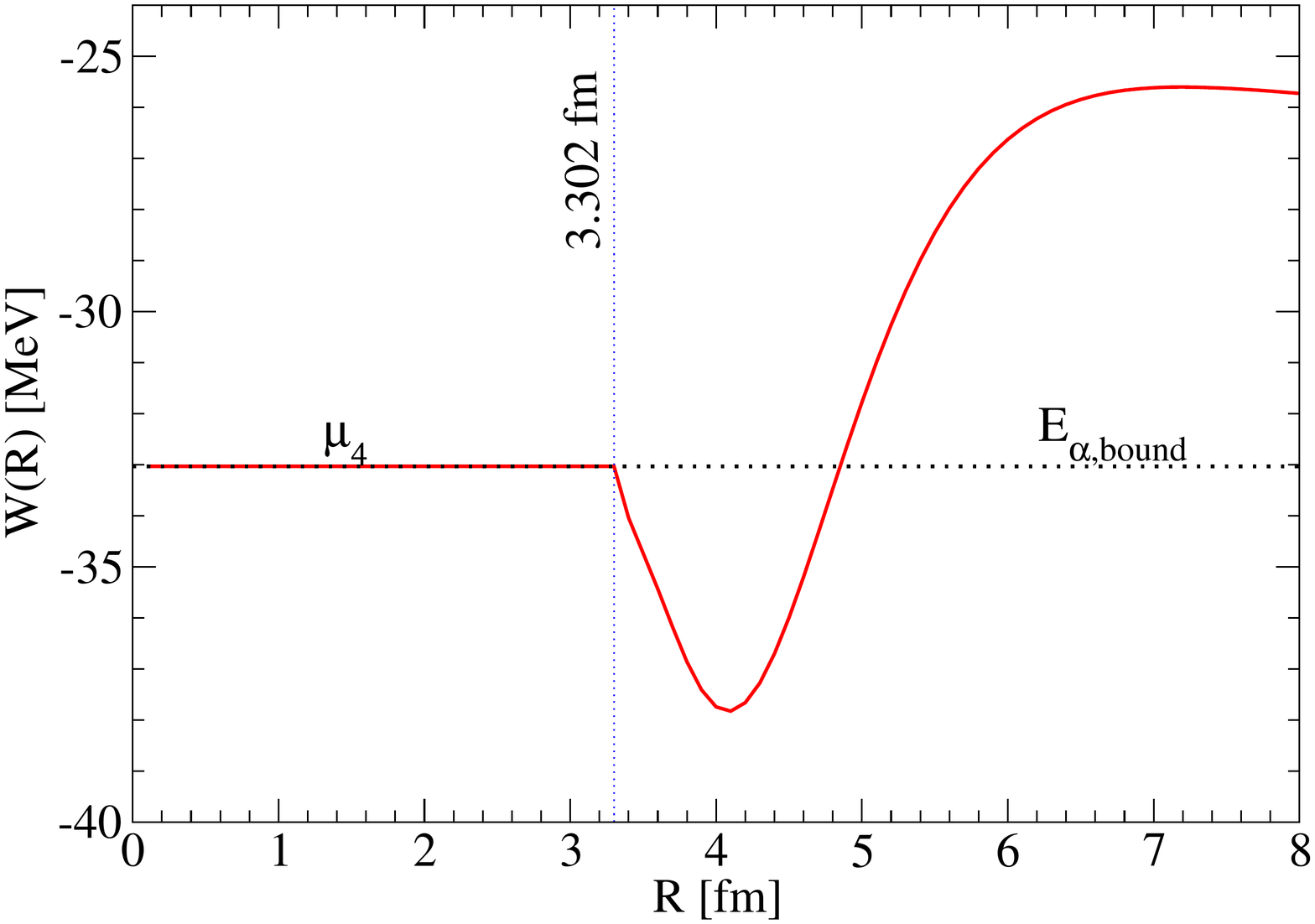}}
  \caption{Effective potential $ W^{\rm TF}( R) $ for the center of mass motion of the quartet on top of $^{16}$O. 
The Thomas-Fermi model has been used. The formation of a pocket is shown.}
\label{fig:2}
\end{figure}


The resulting effective potential $ W^{\rm TF}( R) $ (\ref{WeffR}) 
for the center of mass motion of the quartet is shown in Fig. \ref{fig:2}. 
The formation of a pocket near the surface is seen which is caused by the formation of an $\alpha$-like cluster. The sharp kink at the critical radius 
$r_{\rm crit}=3.302$ fm is a consequence of the local approximation for the Pauli blocking term. A smooth behavior is expected if the finite extension of the 
$\alpha$-like cluster is taken into account so that the kink produced by the local density approximation is smeared out.

\begin{figure}[h]
  \centerline{\includegraphics[width=250pt]{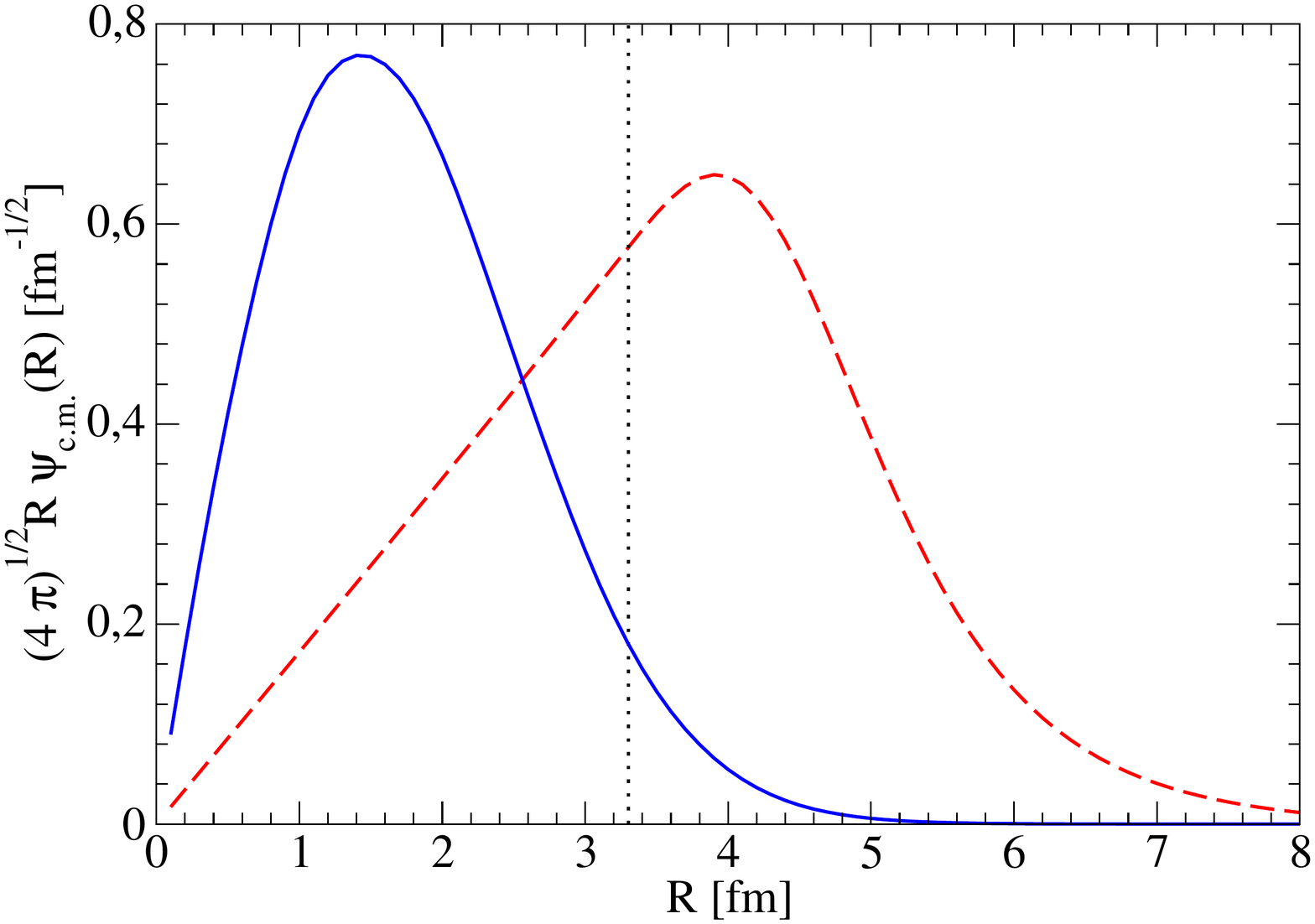}}
  \caption{Wave function for the c.o.m. motion of the quartet. A prefactor $(4 \pi)^{1/2} R$ is introduced so that the integral over $R$ of the squared quantity is normalized to 1. The solution for the Thomas-Fermi model $\psi^{\rm TF}_{\rm c.o.m.}( R)$ (red, dashed) is compared with the non-interacting shell-model calculation $\psi_{2s^4}( R)$ (blue). The shift of the maximum is caused by the formation of a pocket, see Fig.~\ref{fig:2}. Dotted line: $r_{\rm crit}$.}
\label{fig:3}
\end{figure}

The wave function for the quartet center of mass motion $\psi^{\rm TF}_{\rm c.o.m.}( R)$ is found as solution of the Schr\"odinger equation, mass $4 m$, 
with the potential $ W^{\rm TF}( R) $. The energy eigenvalue is -33.0 MeV. A graph of $(4 \pi)^{1/2} R\, \psi^{\rm TF}_{\rm c.o.m.}( R)$ is shown in Fig. \ref{fig:3}. As a result, in Ref. \cite{wir} the rms point radius 2.8644 fm for  $^{20}$Ne
is calculated
which is in good agreement with the experimental rms point radius 2.87 fm. 
%

\section{OPTICAL MODEL DESCRPTION  AND DOUBLE-FOLDING POTENTIAL}

We discuss the effective c.o.m. potential $W^{\rm TF}( R)$ and compare with other approaches. 
In particular we check
whether the choice (\ref{cd}) for the double-folding potential parameter $c,\,\,d$ are realistic.

Elastic scattering of $\alpha$ particles by the $^{16}$O nucleus has been investigated, 
and the corresponding optical potentials have been inferred. There is a large uncertainty 
for small values of $R$.
In a first discussion (McFadden) \cite{Fadden66}, the optical potential, real part
\begin{equation}
\label{optpot}
-\frac{V_0}{1+e^{(r-r_0A^{1/3})/a}}
\end{equation}
with $V_0=43.9$ MeV, $r_0=1.912$ fm and $a=0.451$ fm has been proposed.
A large ambiguity ($V_0=33 - 290$ MeV for Ni) has been observed.

Another approach (Michel) was given in Ref. \cite{Michel} and compared with experiments \cite{Oertzen}
\begin{equation}
-V_0\frac{1+\alpha e^{-(r/\rho)^2}}{[1+e^{(r-R_R)/(2 a_R)}]^2}
\end{equation}
with $V_0=38$ MeV, $\rho = 4.5$ fm, $R_R=4.3$ fm, $a_R=0.6$ fm, and the energy-dependent $\alpha = 3.625$.
More recently, in Ref. \cite{Ohkubo}  a density dependent M3Y effective
interaction (DDM3Y) was used, and a double-folding potential has been derived (Fig. 3 in \cite{Ohkubo}) which goes to -110 MeV at  $R=0$.

In Ref. \cite{Fukui16} (Fukui) a 16O (6Li,d) 20Ne transfer reaction has been considered, where the  model potential (\ref{optpot})
with $r_0=1.25$ fm and $a=0.76$ fm has been used, $V_0$ was adapted to reproduce the value 4.73 MeV of the binding energy. 
Ashok Kumar and S. Kailas \cite{Kumar} give the parameter values $V_0=142.5$ MeV, 
$r_0=1.18$ fm and $a_0=0.76$ fm.


\begin{figure}[h]
  \centerline{\includegraphics[width=250pt]{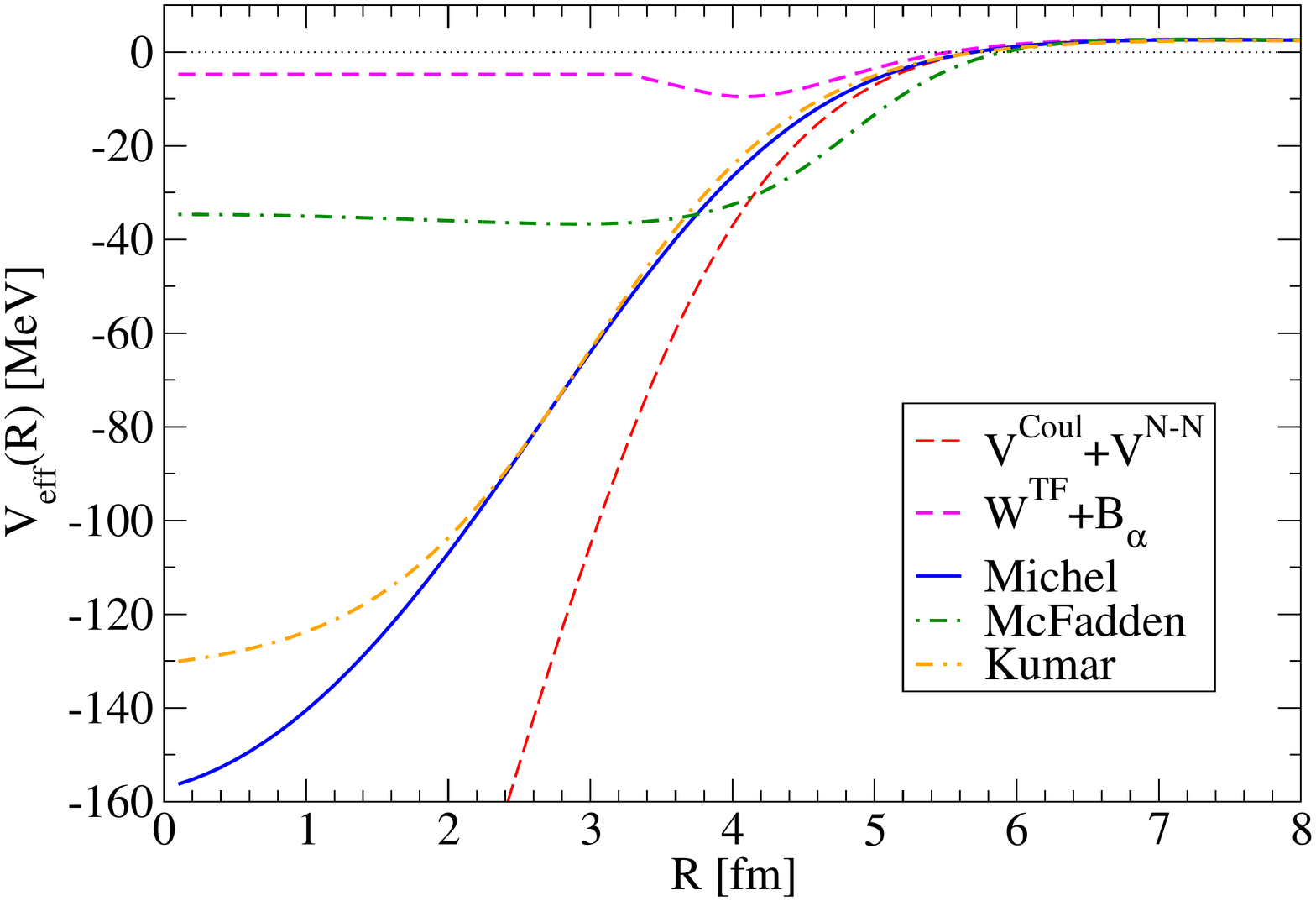}}
  \caption{Optical model potentials (see main text for explanation).}
\label{fig:W}
\end{figure}

Note that $V_{\rm eff}(R)=W( R)+B_\alpha \approx
 V^{\rm Coul}_{\alpha - {\rm O}}( R)+V^{\rm N-N}_{\alpha - {\rm O}}( R)$ 
is the mean field relative to the free $\alpha$ particle.  
Below $R = 5$ fm, Pauli blocking terms arise, see Eq. (\ref{WeffR}).
The coincidence with  \cite{Michel} (Michel) is rather good. We conclude that the choice 
(\ref{cd}) is reasonable.

\section{SHELL MODEL CALCULATION}

The local density approximation (Thomas-Fermi model) is not able to describe the 
nuclear structure of the core nucleus. In particular, the Thomas-Fermi rule has to be replaced 
by a more microscopic approach, see \cite{Po,Xu,Xu1}. However, the behavior of the effective c.o.m. potential 
which remains nearly constant inside the core nucleus is of interest also for the case where 
shell model states are used. A first attempt has been performed in Ref. \cite{wir} 
using harmonic oscillator states. However, the harmonic oscillator potential is not realistic 
for nuclei, in particular near the surface of the core nucleus where $\alpha$-like quartets are formed.
We present here calculations with more realistic potentials (units MeV, fm), see also \cite{Mirea}. The intrinsic nucleon-nucleon interaction 
$W^{\rm intr}( R)$ which is suppressed because of Pauli blocking, is not considered in this Section.

\begin{figure}[h]
  \centerline{\includegraphics[width=240pt]{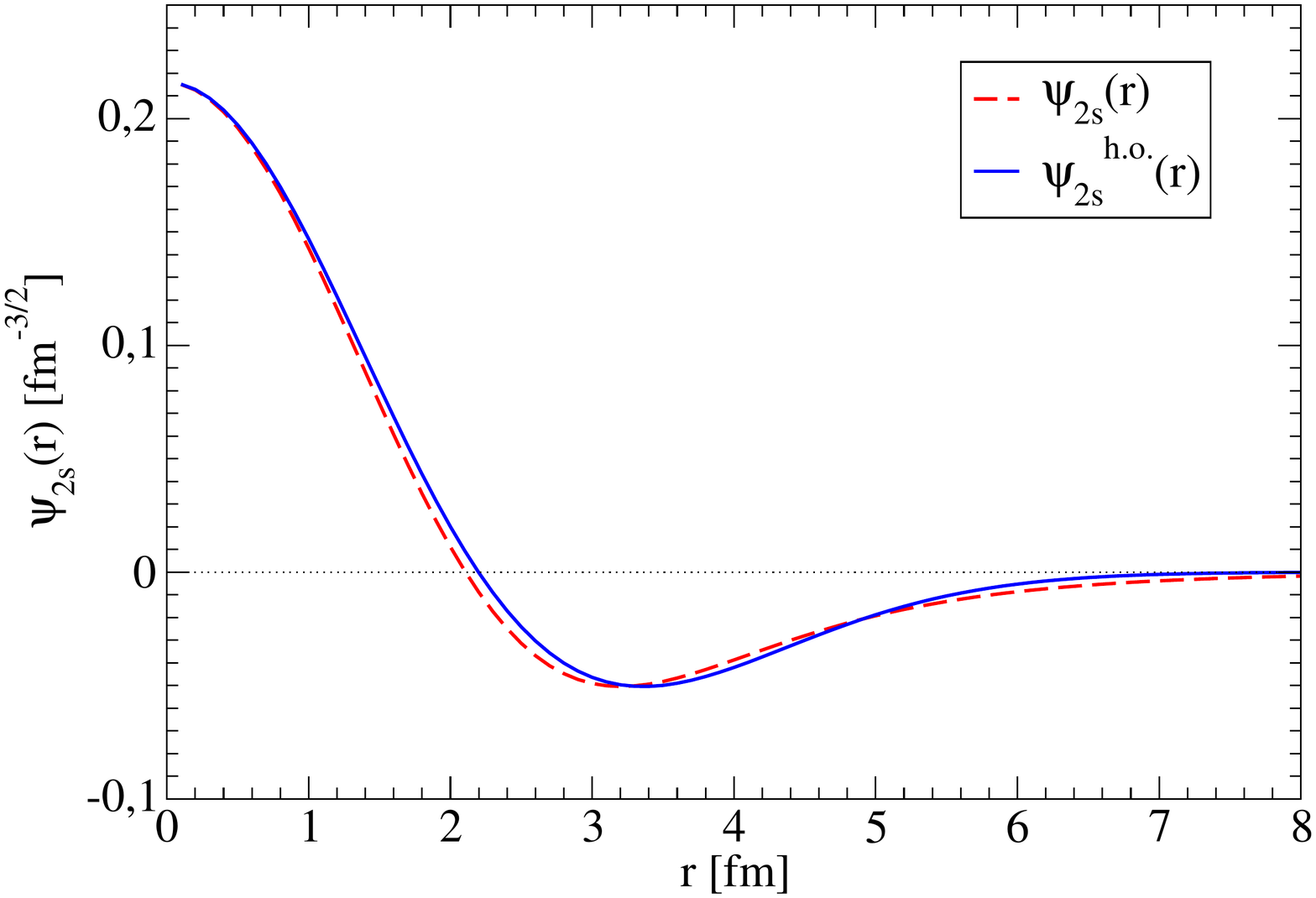}
\includegraphics[width=240pt]{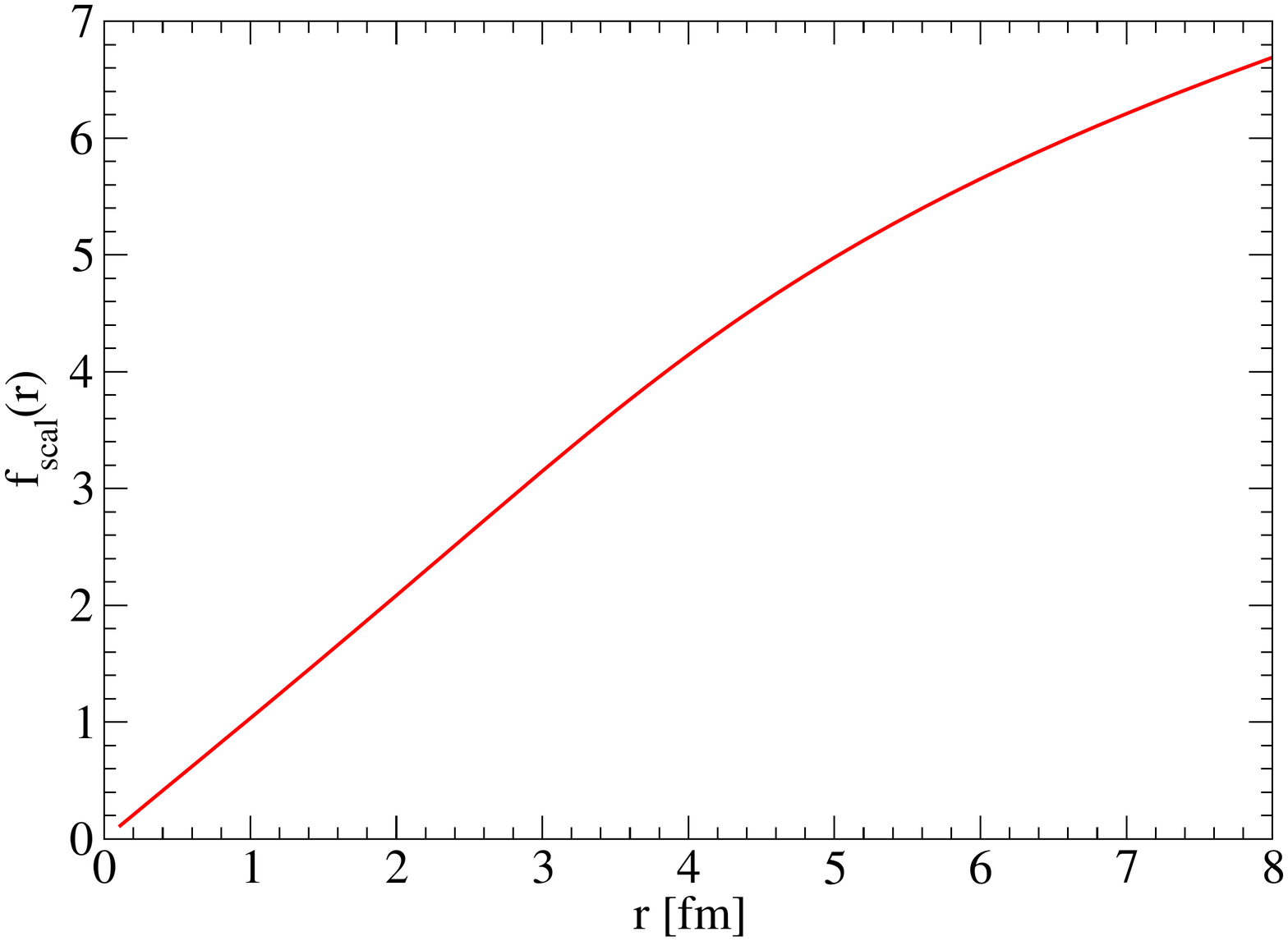}}
  \caption{Normalized wave function $\psi_{2s}(r )$ for the Woods-Saxon potential (\ref{WS2s}). For comparison, the harmonic oscillator wave function $\psi_{2s}^{\rm HO}(r)$ is also given, where the potential parameter $a^{\rm HO}$ is chosen so that 
$\psi_{2s}(0)$ coincides. The scaling function $f_{\rm scal}(r )$ give full coincidence of both wave functions.}
\label{Fig:psi}
\end{figure}

We use a simple description to find an appropriate basis of single-particle states.
We use the Woods-Saxon potential \cite{WS} for $Z=N$
\begin{equation}
\label{WS2s}
 V_{\rm WS}(r)=\frac{V_0 (1+3\kappa/A)}{1+\exp[(r-R_0 A^{1/3})/a]}
\end{equation}
with $V_0=-52.06$ MeV, $\kappa=0.639$, $R_0=1.26$ fm, $a=0.662$ fm. 
The normalized solution $\psi_{2s}(r )$ for the 2$s$ state is shown in Fig. \ref{Fig:psi}, eigenvalue $E_{2s}=-9.162$ MeV. For comparison,
the harmonic oscillator wave function
\begin{equation}
 \psi_{2s}^{\rm HO}(r )=-\left(\frac{a^{\rm HO}}{\pi}\right)^{3/4} e^{-a^{\rm HO} r^2/2}\left(a^{\rm HO} r^2-\frac{3}{2}\right) \left(\frac{2}{3}\right)^{1/2},
\end{equation}
is also shown, where the parameter $a^{\rm HO}=0.31047$ fm is chosen so that 
the values at $r=0$ coincide. A scaling of the $r$-axis is considered to make both coincident, 
$\psi_{2s}^{\rm HO}(r' )=\psi_{2s}(r )/(1+0.0024719\, r)$. (The amplitude correction is necessary to reproduce the correct value of the minimum.)
This defines the relation $r'=f_{\rm scal}(r )$ shown in Fig. \ref{Fig:psi}.

Neglecting any intrinsic interaction, the  2$s$ wave functions can be used to construct the quartet  wave function
\begin{equation}
 \Phi_{2s^4}({\bf R, S, s, s}')=\psi_{2s}({\bf r}_{n,\uparrow})\,\psi_{2s}({\bf r}_{n,\downarrow})\,
\psi_{2s}({\bf r}_{p,\uparrow})\,\psi_{2s}({\bf r}_{p,\downarrow}).
\end{equation}
The wave function for the c.o.m. motion follows as (Jacobi-Moshinsky coordinates ${\bf R, S, s, s}'$ \cite{wir})
\begin{equation}
\label{Phi1s}
\psi_{2s^4}({\bf R})=\left[\int d^3Sd^3sd^3s'|\Phi_{2s^4}({\bf R, S, s, s}')|^2\right]^{1/2}\,.
\end{equation}

The evaluation of the 9-fold integral in (\ref{Phi1s}) is very time-consuming. An approximation can be given comparing with the solution for the harmonic oscillator \cite{wir}
\begin{eqnarray}
\label{rho2s}
&&\varrho_{2s^4}^{\rm cm, HO}(a, R)=|\psi^{\rm HO}_{2s^4}( R)|^2=\left(\frac{a}{\pi}\right)^{3/2} e^{-4a R^2} \frac{1}{10616832} (24695649+14905152\, a R^2\nonumber \\&&
+354818304\, a^2R^4 -876834816\, a^3R^6
+1503289344\, a^4R^8-1261699072\, a^5R^{10} \nonumber\\ &&
+613416960\, a^6R^{12}-150994944\, a^7 R^{14}+16777216\, a^8R^{16}).
\end{eqnarray}
The parameter $a"=0.287038$ fm can be chosen to reproduce the value at $R=0$ (three-fold integral). The scaling $R"=f_{\rm scal}(R) + 0.174 \,(e^{R/2.924}-1)$ fulfills normalization and improves the asymptotic behavior for large $R$, so that $\varrho_{2s^4}^{\rm cm}(R)\approx \varrho_{2s^4}^{\rm cm, HO}(a", R")$. 
A plot of $(4 \pi R^2)^{1/2} \psi_{2s^4}( R)$ is shown in Fig. \ref{fig:3}. The normalization $\int_0^\infty 4 \pi R^2 \psi^2_{2s^4}( R) dR =1$ holds.

\begin{figure}[h]
  \centerline{\includegraphics[width=250pt]{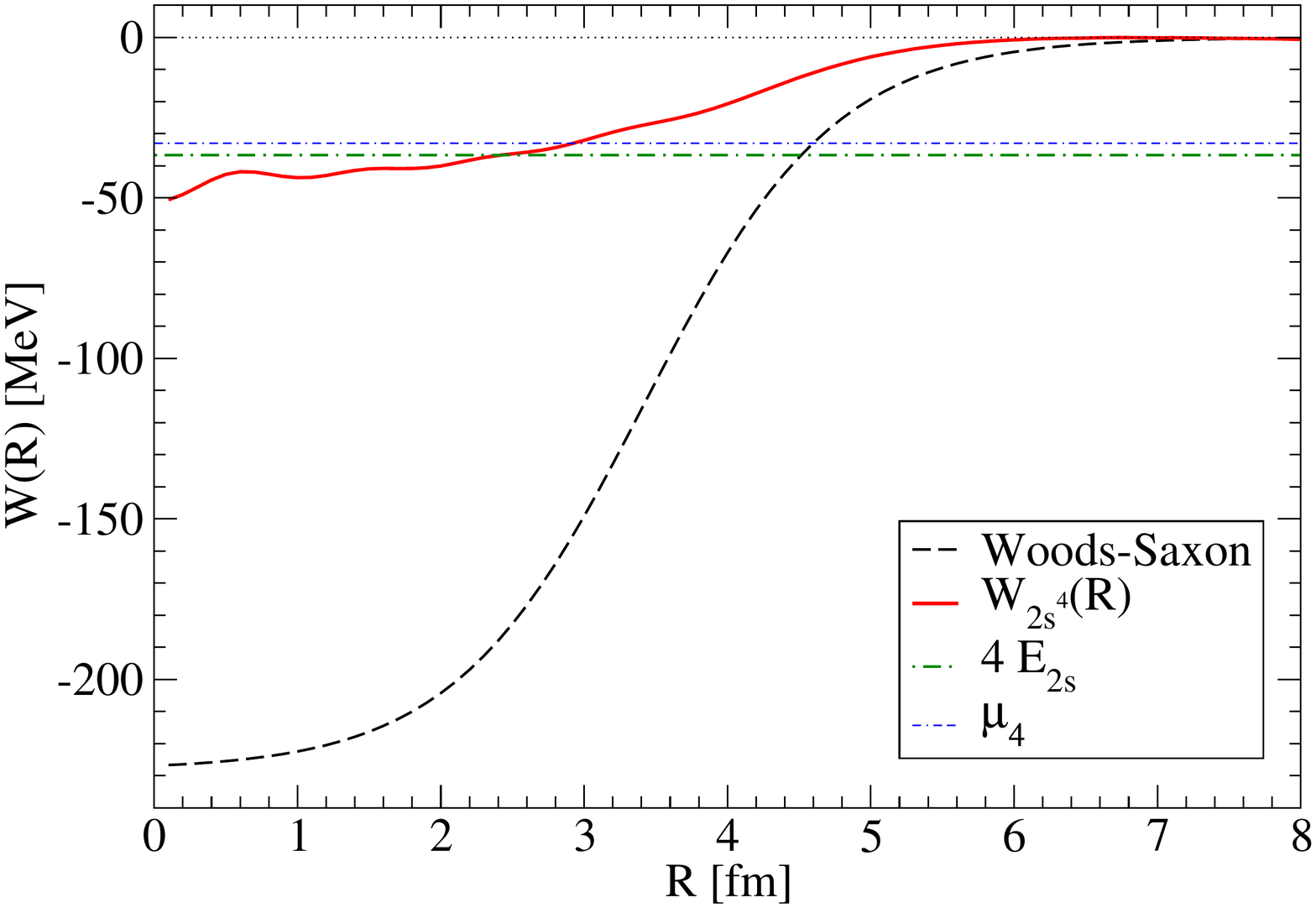}}
  \caption{The c.o.m. potential $W_{2s^4}( R)$, Eq. (\ref{Sglcom1}), compared with the Woods-Saxon potential of the quartet.}
\label{Fig:W}
\end{figure}

We reconstruct the effective potential from the wave function  
$\psi_{2s^4}( R)=(\varrho_{2s^4}^{\rm cm}( R) )^{1/2}$  \cite{wir}.
If we restrict us to $s$ states ($l=0$) and introduce $u_{2s^4}( R)= (4 \pi)^{1/2} R \psi_{2s^4}( R)$, 
we have
\begin{equation}
\label{Sglcom1}
W_{2s^4}( R)-E_{2s^4}=\frac{\hbar^2}{8m} \frac{1}{u_{2s^4}( R)} \frac{d^2}{dR^2} u_{2s^4}( R).
\end{equation}
The result is shown in Fig. \ref{Fig:W}.

We conclude: The effective c.o.m. potential $W( R)$ remains nearly constant within the core 
as expected from the Thomas-Fermi model. The value $E_{2s^4} = -36.65$ MeV 
is near to the estimate $\mu_4=-33$ MeV from the Thomas-Fermi rule.
It is slightly increasing near the surface, possibly because the quartet is not point-like localized, 
but smeared out so that it "feels" the weakening of the potential near the surface. Another reason may be the gradient terms in Eq. (\ref{9}) which are neglected here.
A similar behavior has also been observed for the harmonic oscillator potential in \cite{wir}.
In contrast to the harmonic oscillator where the effective potential increases with $R$, 
now the behavior near the surface is more realistic.
Relaxing the Thomas-Fermi rule has been discussed in Refs. \cite{Po,Xu,Xu1,wir}.

\section{INTRINSIC INTERACTION AND PAULI BLOCKING}

\begin{figure}[h]
  \centerline{\includegraphics[width=250pt]{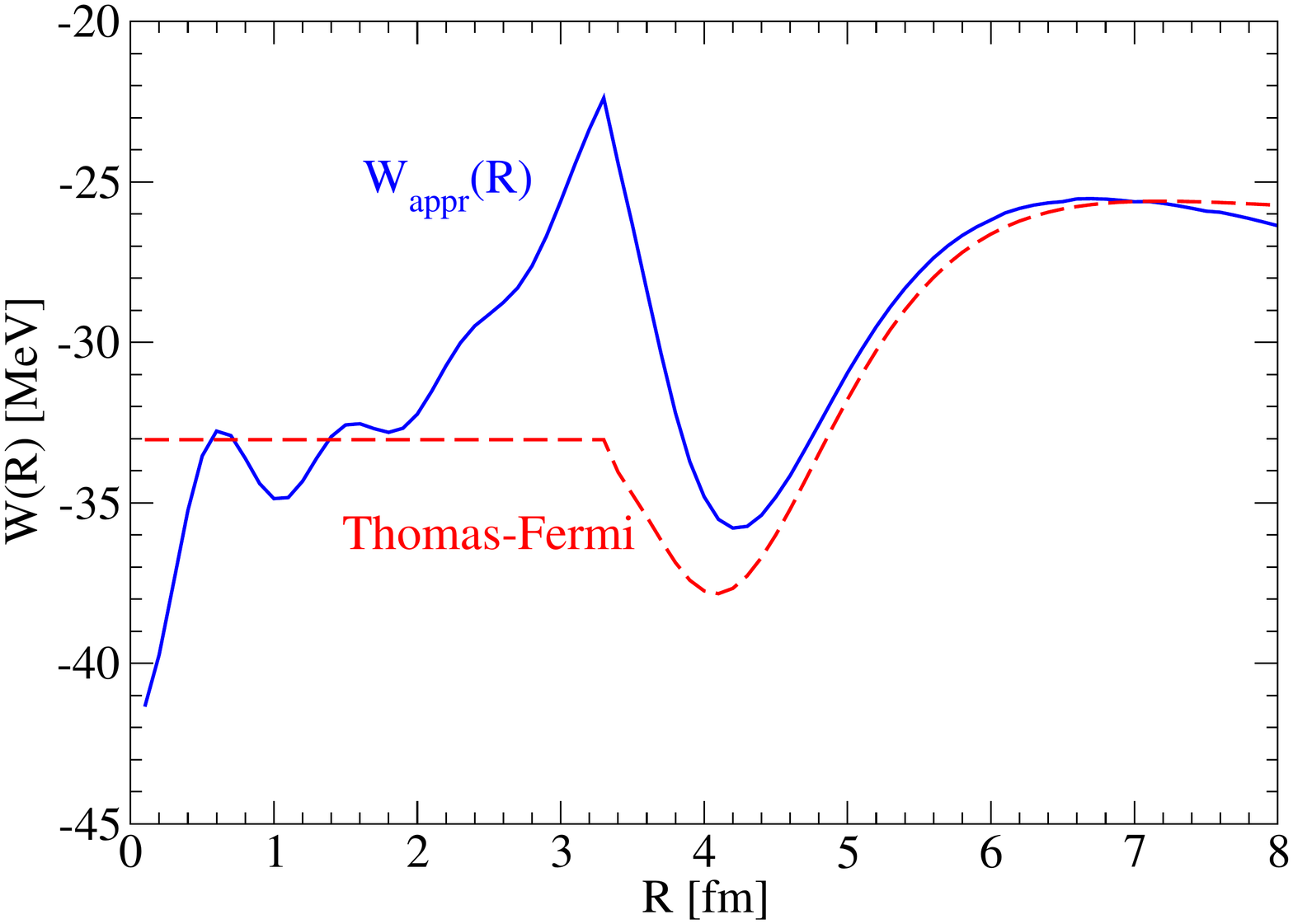}}
  \caption{Quartet c.o.m. potentials $W(R )$. The Thomas-Fermi approximation $W^{\rm TF}(R )$ 
is compared with the calculation $W_{\rm appr}(R )$ using shell model states.}
\label{Fig:pocket}
\end{figure}

To give an estimate of the intrinsic interaction of the quartet we add for $R > r_{\rm crit}$ 
the formation of the cluster and dissolution because of Pauli blocking as given in Eq. (\ref{WeffR}), 
see Fig. \ref{Fig:pocket}. The Coulomb potential is added, and the shell-model free effective potential 
$W_{2s^4}( R)$ is used instead of $W^{\rm ext}$. Obviously this c.o.m. potential $W_{\rm appr}(R )$ 
is only a rough approximation 
in the low-density region where clusters may be formed. Quartet correlations are present also for 
$R \le r_{\rm crit}$,  and the large peak near $R=3$ fm may be obsolete. Because in the local density 
approximation the dissolution of the cluster happens sharply at the critical radius, there is a sharp 
effect which produces the peak. It is expected that the behavior remains smooth if also the blocking term 
is smeared out, improving the local density approximation (\ref{WeffR}). 
Note the good coincidence with $W^{\rm TF}(R )$
for  $R > r_{\rm crit}$ which is based on the double folding potential, in contrast to the pure nucleonic potential $4  V_{\rm WS}(r)$, Eq. (\ref{WS2s}).

\section{CONCLUSIONS, COMPARISON WITH THE THSR MODEL}

We investigated the c.o.m. motion of an $\alpha$-like quartet moving under the influence of 
a core nucleus, here the $^{16}$O nucleus. In local density approximation, 
an effective potential $W( R)$ for the quartet c.o.m. motion 
has been obtained, which shows a pocket structure near the surface of the nucleus 
what is of relevance for the preformation of $\alpha$ particles. 
A new aspect is the behavior of  $W( R)$ within the core nucleus, 
i.e. for $R \le r_{\rm crit}$ where the quartet bound state is dissolved
because of Pauli blocking. 
In contrast to former investigations which assume an increase of the
effective $\alpha- ^{16}$O potential with decreasing $R$, within a Thomas-Fermi approach  
$W^{\rm TF}(R)=\mu_4$ remains constant in this region $R \le r_{\rm crit}$ \cite{Po,Xu,Xu1,wir}, 
see Fig. \ref{fig:2}. 
In the present work we show also for
the shell model approach, that the effective potential $W(R )$ remains nearly constant 
inside the core nucleus. The reason is the exchange interaction or the Pauli blocking 
between the quartet nucleons and the core nucleus.

For large distances, the empirically determined M3Y potential used for  $W(R )$ (parameter $c, d$) 
seems to fit well
with the optical potentials deduced from scattering experiments. Near the surface of the nucleus, 
Pauli blocking becomes relevant. A pocket is formed for the effective potential $W^{\rm TF}(R)$ and 
remains also after introducing single-particle shell model states for the core nucleus. 
However, the local density approximation for the Pauli blocking should be improved,
and it is expected that sharp peak structures are smeared out.

Of interest is the comparison with the THSR approach \cite{THSR,Toh17} which treats the quartets 
self-consistently. If for the surrounding medium a mean-field description based on 
uncorrelated single-particle states is no longer possible, correlations in the medium,
in particular quartetting, should be considered. The full antisymmetrization 
of the many-body wave function is very challenging. The THSR
approach provides us with such a self-consistent, antisymmetrized treatment of quartetting of all nucleons. 
A variational principle with Gaussian wave functions has been used, and nuclei with $A \le 20$
have been treated this way. Interesting results have been obtained for $^{20}$Ne \cite{Bo,Bo2,Bo3} 
considering the full antisymmetrization of $\alpha$ and $^{16}$O wave functions 
using Gaussian distributions.
However, it is not trivial to find the appropriate observables in the THSR approach to deduce an effective potential $W( R)$ and wave function $\psi( R)$ for the quartet c.o.m. motion.

The comparison of our quartetting wave function approach using shell-model states for the core nucleus,
with the THSR approach may answer the question whether quartetting is also relevant for the core nucleus.
In our present approximation, the core nucleus is described by uncorrelated, single-nucleon states.
It is known \cite{THSR} that $\alpha$-like correlations are also present 
in the ground state of the $^{16}$O core nucleus. Within THSR calculations, the minimum of the energy functional has been found for Gaussians
with different width $b, B$ for the intrinsic cluster wave function 
and the c.o.m. wave function, respectively, see  Fig. 2 of Ref. \cite{THSR}. 
This indicates that the assumption of a shell model of independent single-nucleon orbits 
is not fully justified for the $^{16}$O core nucleus.
The THSR approach gives a self-consistent approach to quartetting since quartetting may occur 
also in the surrounding nuclear medium, that not always is a compact core nucleus. 
We refer to the Hoyle state in $^{12}$C as example.
However, the restriction to only Gaussian-type wave functions can be improved.

\section{ACKNOWLEDGMENTS}
I thank my colleagues for close collaboration in this field: P. Schuck from France, as well as  H. Horiuchi,  A. Tohsaki, Y. Funaki, T. Yamada from Japan, and  Z. Ren, C. Xu, Bo Zhou, M. Lyu,
Q. Zhao from China.

\end{document}